\documentclass[runningheads, anonymous]{llncs}
\usepackage{listings}
\usepackage{amssymb}
\usepackage{amsmath}
\usepackage{cite}
\usepackage{graphicx}
\usepackage{comment}
\usepackage[noline,boxed,ruled,noend,linesnumbered]{algorithm2e}
\usepackage{wrapfig}
\usepackage{vcell}
\usepackage{appendix}
\usepackage[hidelinks]{hyperref}
\usepackage{lmodern}
\usepackage{booktabs}
\usepackage{makecell}
\usepackage{multirow}
\usepackage{pifont} 
\usepackage[usenames,dvipsnames]{xcolor}

\usepackage{hyperref}
\hypersetup{
    colorlinks=true,
    linkcolor=RedViolet,
    anchorcolor=black,
    citecolor=blue,
    urlcolor=black,
    breaklinks=true
}

\definecolor{coqkeyword}{RGB}{128, 0, 128}      
\definecolor{coqtactic}{RGB}{175, 135, 0}       
\definecolor{coqcomment}{RGB}{34, 139, 34}      

\lstdefinestyle{coqstyle}{
  language=rocq,
  basicstyle=\ttfamily\small,
  keywordstyle=[1]\color{coqkeyword}\bfseries,  
  keywordstyle=[2]\color{coqtactic},             
  commentstyle=\color{coqcomment},               
  showstringspaces=false,
  breaklines=true,
  columns=flexible,
  keepspaces=true,
  xleftmargin=1em,
  frame=none,
}

\usepackage[capitalise]{cleveref}
\Crefname{equation}{}{}
\Crefname{appendix}{Appx.}{Appxs.}
\Crefname{section}{Sect.}{Sects.}
\Crefname{figure}{Fig.}{Figs.}

\usepackage[inline]{enumitem}
\usepackage{csquotes}
\usepackage{marvosym}
\usepackage{multirow}
\usepackage{color}
\usepackage[T1]{fontenc}
\input{macro}

\newcommand{\toolname}[0]{{\sc ReCent-Prover}}

\newlist{compitem}{itemize}{4}
\setlist[compitem,1]{nolistsep,label=$\bullet$}

\newenvironment{talign*}
 {\let\displaystyle\textstyle\csname align*\endcsname}
 {\endalign}

\makeatletter 
\@mparswitchfalse%
\makeatother
\normalmarginpar 

\begin{document}

\title{On Reasoning-Centric LLM-based Automated Theorem Proving}

\author{}
\authorrunning{}
\institute{}

\author{Yican Sun\inst{1} \and
Chengwei Shi \inst{1}  \and
Hangzhou Lyu \inst{1}\and
Yingfei Xiong \inst{1}\inst{2}$^{\text{(\Letter)}}$
}

\authorrunning{Y.~Sun et al.}
\institute{Key Laboratory of High Confidence Software Technologies (Peking University), Ministry of Education; School of
Computer Science, Peking University, Beijing, China\\ \email{\{sycpku,xiongyf\}@pku.edu.cn}\\\email{\{2200013126,2300012939\}@stu.pku.edu.cn}\and
Zhongguancun Laboratory 
}

\maketitle             
\begin{abstract}
Automated theorem proving is fundamental to formal methods, and the recent trend is to integrate large language models (LLMs) and proof assistants to form effective proof agents. While existing proof agents show promising performance, they inadequately leverage reasoning capabilities of modern LLMs in high-level planning and self-critique. We argue that proof agents should not merely generate tactics but also reason strategically about proof plans and critically evaluate their own proposals. 

This paper introduces \toolname{}, a reasoning-centric LLM-based proof agent for Rocq that addresses two critical limitations in current systems. First, we present \emph{validation with reflection}, enabling LLMs to scrutinize their generated tactics and synthesize failure summaries when reflection identifies potential errors, filtering out potentially misapplied tactics earlier. Second, we propose \emph{retrieval with planning}, which conditions retrieval on LLM-generated proof plans rather than subgoal similarity, retrieving lemmas and proofs that align with the anticipated proof strategy. Both techniques increase the number of invocations of LLMs. However, when evaluated on the CoqStoq benchmark, even under the same budget of LLM invocations, \toolname{} achieves a 22.58\% relative improvement in the number of proved theorems over the previous state-of-the-art, demonstrating that our reasoning-centric design significantly enhances automated theorem proving capabilities.
\end{abstract}

\newcommand\code[1]{{\tt\small #1}}
\definecolor{dkgreen}{rgb}{0,0.3,0}
\definecolor{cmtgreen}{rgb}{0,0.6,0}
\definecolor{ltblue}{rgb}{0,0.4,0.4}
\definecolor{dkviolet}{rgb}{0.3,0,0.5}
\definecolor{dkblue}{rgb}{0,0.2,0.2}
\definecolor{dkred}{rgb}{0.6,0,0}
\lstdefinelanguage{rocq}{ 
    mathescape=true,
    texcl=false, 
    escapeinside={(@}{@)},
    morekeywords=[1]{Section, Module, End, Require, Import, Export,
        Variable, Variables, Parameter, Parameters, Axiom, Hypothesis,
        Hypotheses, Notation, Local, Tactic, Reserved, Scope, Open, Close,
        Bind, Delimit, Definition, Let, Ltac, Fixpoint, CoFixpoint, Add,
        Morphism, Relation, Implicit, Arguments, Unset, Contextual,
        Strict, Prenex, Implicits, Inductive, CoInductive, Record,
        Structure, Canonical, Coercion, Context, Class, Global, Instance,
        Program, Infix, Theorem, Lemma, Corollary, Proposition, Fact,
        Remark, Example, Proof, Goal, Save, Qed, Defined, Hint, Resolve,
        Rewrite, View, Search, Show, Print, Printing, All, Eval, Check,
        Projections, inside, outside, Def},
    morekeywords=[2]{forall, exists, exists2, fun, fix, cofix, struct,
        match, with, end, as, in, return, let, if, is, then, else, for, of,
        nosimpl, when},
    morekeywords=[3]{Type, Prop, Set, true, false, option},
    morekeywords=[4]{pose, set, move, case, elim, apply, clear, hnf,
        intro, intros, generalize, rename, pattern, after, destruct,
        induction, using, refine, inversion, injection, rewrite, congr,
        unlock, compute, ring, field, fourier, replace, fold, unfold,
        change, cutrewrite, simpl, have, suff, wlog, suffices, without,
        loss, nat_norm, assert, cut, trivial, revert, bool_congr, nat_congr,
        symmetry, transitivity, auto, split, left, right, autorewrite},
    morekeywords=[5]{by, done, exact, tauto, romega, omega,
        assumption, solve, contradiction, discriminate},
    morekeywords=[6]{do, last, first, try, idtac, repeat},
    morecomment=[s]{(*}{*)},
    showstringspaces=false,
    morestring=[b]",
    morestring=[d]’,
    tabsize=3,
    extendedchars=false,
    sensitive=true,
    breaklines=false,
    basicstyle=\scriptsize\ttfamily,
    captionpos=b,
    columns=[l]flexible,
    identifierstyle={\ttfamily\color{black}},
    keywordstyle=[1]{\bfseries\ttfamily\color{dkviolet}},
    keywordstyle=[2]{\bfseries\ttfamily\color{dkgreen}},
    keywordstyle=[3]{\bfseries\ttfamily\color{ltblue}},
    keywordstyle=[4]{\bfseries\ttfamily\color{dkblue}},
    keywordstyle=[5]{\bfseries\ttfamily\color{dkred}},
    stringstyle=\ttfamily,
    commentstyle={\bfseries\ttfamily\color{cmtgreen}},
    literate=
    {\\forall}{{\color{dkgreen}{$\forall\;$}}}1
    {\\exists}{{$\exists\;$}}1
    {<-}{{$\leftarrow\;$}}1
    {=>}{{$\Rightarrow\;$}}1
    {==}{{\code{==}\;}}1
    {==>}{{\code{==>}\;}}1
    {->}{{$\rightarrow\;$}}1
    {<->}{{$\leftrightarrow\;$}}1
    {<==}{{$\leq\;$}}1
    {\#}{{$^\star$}}1 
    {\\o}{{$\circ\;$}}1 
    {\@}{{$\cdot$}}1 
    {\/\\}{{$\wedge\;$}}1
    {\\\/}{{$\vee\;$}}1
    {++}{{\code{++}}}1
    {~}{{$\sim$}}1
    {\@\@}{{$@$}}1
    {\\mapsto}{{$\mapsto\;$}}1
    {\\hline}{{\rule{\linewidth}{0.5pt}}}1
}[keywords,comments,strings]

\renewcommand{\paragraph}[1]{\medskip\noindent\textbf{{{#1.}}}}

\section{Introduction}
\label{sec:intro}

Theorem proving within proof assistants is the cornerstone of formal methods, enabling landmark achievements from verifying compilers~\cite{compcert} to mechanizing the proofs of mathematical theorems~\cite{coq-fourcolor,coq-coqeal,coq-reglang}. In the era of generative AI, the need to formally verify AI-generated artifacts---including both code and mathematical proofs---makes theorem proving even more critical. However, constructing formal proofs is labor-intensive. Traditional automation techniques based on symbolic proof search~\cite{coq-hammer,lean-auto,sledgehammer,norman2025canonical} have provided valuable assistance, but they fundamentally struggle to scale to complex tasks due to the explosion of the search space.

The recent success of large language models (LLMs) has reshaped automated theorem proving~\cite{badlur,rango,cobblestone,copra,PALM,lean-dojo,lenabel-prover,dsprover,goedel-prover,DSP,ERP}, shifting the paradigm from exhaustive symbolic search to neural proof generation. Since LLMs rarely generate correct proofs in a single attempt, state-of-the-art systems~\cite{cobblestone} adopt an agentic approach where an LLM iteratively generates formalized proofs in an interactive environment of a proof assistant. To help the LLM better understand the current subgoal, these systems retrieve relevant lemmas and definitions to contextualize the proof task, with some~\cite{rango} further incorporating existing proofs to guide the proof strategies.

Despite substantial progress, existing systems largely emphasize the mechanics of interaction between LLMs and proof assistants (e.g., how to incorporate the feedback from the proof assistant), rather than fully harnessing the LLMs' capability in high-level planning and self-critique to guide the proof search itself. As LLMs continue to advance, especially the advancement of \emph{reasoning abilities}, we advocate that LLM-based self-reasoning should be the central component of future proof agents. Thus, agents should not only generate tactics, but also {reason} strategically about proof plans and critically evaluate their own proposals. From this perspective, we identify two key limitations of current approaches and propose novel modules to address these limitations.

First, existing proof agents accept every generated tactic validated by the proof assistant, even when it irreversibly makes the goal more difficult to prove or directs the search toward unprovable branches. For instance, misapplying a lemma via \texttt{apply} may generate subgoals that have counter-examples, while inducting on the wrong variable or failing to properly generalize the hypothesis in \texttt{induction} can be ineffective, or even lead to unprovable subgoals.

To address this problem, our solution is to leverage the reasoning capabilities of LLMs to self-validate its own generated tactics.
In detail, we introduce \emph{validation with reflection}, a technique that not only validates LLM-generated tactics using the proof assistant, but also enables the LLM to scrutinize its own tactic proposals. When reflection identifies a potentially misapplied tactic, the agent synthesizes and records a failure summary to steer the next iteration; otherwise, the tactic proceeds as usual. Since reflecting on every tactic would be expensive, we adopt a selective strategy, applying self-reflection only to tactics that may (1) generate more than one subgoals, or (2) lead to unprovable states, such as \texttt{apply} and \texttt{induction} mentioned above.

The second limitation concerns the retrieval process. 
Ideally, a retriever would return lemmas that are actually used in the proof and proofs that instantiate the same high-level strategy. However, this oracle behavior is unattainable without knowing the proof in advance. Existing retrieval methods therefore rely on a proxy assumption---that similar subgoals imply similar proof plans and lemma usage---which does not necessarily hold.

To address this, we observe that although we cannot know the final proof in advance, we can leverage the high-level proof plan, which LLMs excel at generating, as a proxy for retrieval. We therefore propose \emph{retrieval with planning}. We pre-process the lemma library by generating a natural-language description for each lemma using an LLM, then index lemmas by semantic embeddings of these descriptions rather than by their formal statements. Similarly, for each proof example, we generate a natural-language proof plan and index the example database by embeddings of these plans rather than by subgoals. During proof search, we first prompt the LLM to generate a proof plan for the current goal, then use this plan to retrieve candidate lemmas and proofs that align with the anticipated strategy. By conditioning retrieval on plan-level semantics, our approach captures strategies that transfer across syntactically different theorems and surfaces genuinely relevant knowledge beyond subgoal similarity.

We implement and evaluate \toolname{}\footnote{Short for ``Reasoning-Centric Prover''}, a Rocq proof agent integrating validation with reflection and retrieval with planning, on CoqStoq~\cite{rango}, a benchmark of real-world Rocq projects. Despite requiring more LLM invocations per proof attempt, under the same budget of LLM invocations, \toolname{} achieves a 22.58\% relative improvement in proved theorems over the previous state-of-the-art~\cite{cobblestone}. Further experiments over different configurations of \toolname{} demonstrate the effectiveness of both techniques.

\paragraph{Contributions}
To summarize, this paper makes the following contributions:
\begin{itemize}
  \item We propose a novel validation with reflection technique that enables the LLM to critique its own generated tactics during proof search, making the proof search more robust and effective.
  \item We propose a novel retrieval with planning technique that conditions retrieval on LLM-generated proof plans, enabling strategy-aligned lemma and example selection beyond subgoal similarity.
  \item We implement \toolname{}, a Rocq proof agent that integrates two techniques mentioned above, and evaluate it on the CoqStoq benchmark. Our approach achieves a 22.58\% relative improvement in the number of proved theorems against the previous state-of-the-art LLM-based proof agent.
\end{itemize}

\section{Preliminaries}
\label{sec:prelims}

This section introduces basic concepts of formal theorem proving in Rocq (formerly known as Coq)~\cite{CoqTool} through the following running example.

\begin{example}
\label{ex:running-start}
Suppose one is interested in proving the following lemma from CompCert's memory-initialization routine, which characterizes alignment checking for a concatenation of data items laid out consecutively in memory. The theorem statement and relevant definition are presented in \Cref{fig:running-example}. Here, \texttt{data\_list\_size} computes the total size of a data list, and \texttt{dl\_align p il} (presented below) asserts that the list \texttt{il} satisfies the alignment constraints when laid out consecutively starting at position \texttt{p}. The lemma \texttt{dl\_align\_app} reduces alignment of the concatenation to two independent checks: \texttt{l1} at \texttt{pos} and \texttt{l2} at the shifted position \texttt{pos + data\_list\_size l1}. 

\begin{figure}
\begin{lstlisting}[language=rocq]
Theorem dl_align_app: forall l2 l1 pos,
  dl_align pos (l1 $+\hspace{-0.5em}+$ l2) <->
  dl_align pos l1 /\ dl_align (pos + data_list_size l1) l2.

(* the size of an individual datum *)
Definition data_size (i: datum) : Z := ...
(* the sum of data_size of list elements *)
Definition data_list_size (l: list datum) : Z := ...
(* well_aligned i p: datum i is correctly aligned at position p. *)
Definition well_aligned (i: datum) (p: Z) : Prop := ... (* omitted *)
Fixpoint dl_align (p: Z) (il: list datum) : Prop :=
  match il with
  | nil => True
  | i1 :: il => well_aligned i1 p /\ dl_align (p + data_size i1) il
  end.
\end{lstlisting}
\caption{Running Example}
\label{fig:running-example}
\end{figure}
\end{example}

\paragraph{Proof Library} A \emph{proof library} is a collection of (i) previously established lemmas imported at the point where the proof of \texttt{dl\_align\_app} begins, and (ii) the proofs of these imported lemmas. These lemmas can be reused to significantly simplify the proof. Moreover, existing proofs in the library may provide useful proof patterns or insights that can be adapted when constructing the formal proof of \texttt{dl\_align\_app}.

\paragraph{Subgoals} Once the proof is initiated, the proof assistant returns to our prover a list of \emph{subgoals} to be established. Each subgoal consists of a set of premises and a consequent, requiring the prover to prove the consequent from the premises. 

\begin{example}
\label{ex:subgoal}   
Continuing with \Cref{ex:running-start}, at the start of the proof of \texttt{dl\_align\_app}, there is exactly one subgoal presented below.
\begin{lstlisting}[language=rocq,xleftmargin=3em,escapechar=|,xrightmargin=4em]
[No Premise]
|\hrulefill|
forall (l1 l2 : list datum) (pos : Z),
dl_align pos (l1 $+\hspace{-0.5em}+$ l2) <->
dl_align pos l1 /\ dl_align (pos + data_list_size l1) l2
\end{lstlisting}
This state has no premise, and its goal is precisely the statement of \texttt{dl\_align\_app}, meaning that we aim to prove the theorem without any additional assumptions.
\end{example}

\paragraph{Tactics} Subgoals are proved or transformed by applying \emph{tactics}, each of which consumes a subgoal and produces zero, one, or more new subgoals (with zero indicating that the subgoal is proved). Tactics are applied sequentially, one at a time. Each tactic application is checked by the proof assistant. Applying a syntactically valid tactic produces an updated list of subgoals, whereas applying an invalid tactic results in an error and leaves the current subgoals unchanged. By default, tactics are applied to the \emph{first subgoal} in the remaining subgoal list, and all subgoals produced are \emph{prepended} as the first element to the list of unproved subgoals.
\begin{example}
\label{ex:application}
Continuing with \Cref{ex:subgoal}, suppose we apply the tactic \texttt{intros l1 l2 pos.} This introduces \texttt{l1}, \texttt{l2}, and \texttt{pos} into the context, yielding the following subgoal:
\begin{lstlisting}[language=rocq,xleftmargin=3em,escapechar=|,xrightmargin=4em]
l1, l2: list datum
pos: Z
|\hrulefill|
dl_align pos (l1 $+\hspace{-0.5em}+$ l2) <->
dl_align pos l1 /\ dl_align (pos + data_list_size l1) l2
\end{lstlisting}    
\end{example}

Having introduced the relevant concepts, the objective of formal theorem proving is to find a sequence of tactics that (1) can be successfully checked by the proof assistant, and (2) leaves no remaining subgoals after all tactics have been applied. 
\section{The Workflow of \toolname{}}
\label{sec:approach}

\begin{figure}[htbp]
    \centering
    \includegraphics[width=0.8\textwidth]{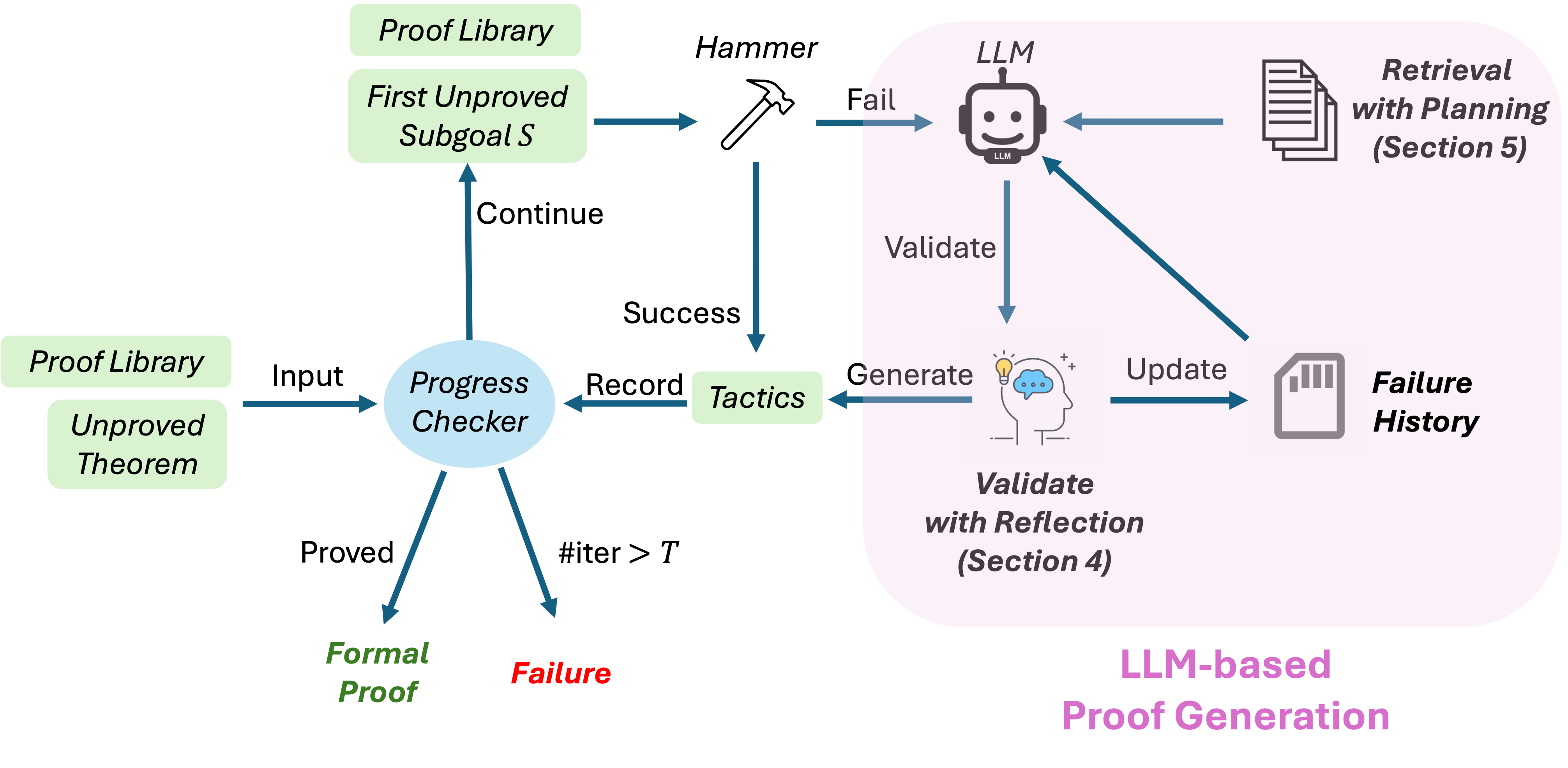}
    \caption{Workflow of \toolname{}.}
    \label{fig:workflow}
\end{figure}

In this section, we present the workflow of \toolname{} (illustrated in \Cref{fig:workflow}) by continuing the running example introduced in (\Cref{ex:running-start}). 

\toolname{} proceeds iteratively. The prover first checks whether there are any remaining subgoals. If no subgoals remain, the prover reports success and returns a complete formal proof. Otherwise, if the iteration limit~$T$ has been reached, the prover reports failure. If neither condition holds, the prover continues proving. When continuing, \toolname{} attempts to prove the first element $S$ among the remaining subgoals, which is the default target for subsequent tactic applications. For instance, the first iteration for proving the theorem in \Cref{ex:running-start} will try to solve the subgoal in \Cref{ex:subgoal}.
\paragraph{Invoking Hammer} To prove $S$, 
our prover first tries to invoke CoqHammer~\cite{coq-hammer}, a state-of-the-art symbolic prover for Rocq. CoqHammer performs symbolic proof search over $S$, automatically selecting and reusing previously established lemmas from the proof library via its built-in lemma selection module.
If CoqHammer succeeds, it generates a formal proof that proves $S$. In this case, our prover records the generated proof and advances to the next iteration. If CoqHammer fails, the prover falls back to invoking an LLM to generate tactics. 

\paragraph{Retrieval with Planning}
To guide the LLM in generating tactics, we need to provide the LLM not only with the current subgoal $S$ but also with essential information in the proof library to make the LLM aware of existing lemmas and proofs. However, including the entire proof library is impractical due to the input-length limitations of LLMs. Thus, it is necessary to retrieve a subset of the library that is most useful for proving the current subgoal. \toolname{} therefore performs lemma and proof-example retrieval, taking the current subgoal \(S\) as input and producing two outputs: (1) a set of lemmas that are expected to be reused in the proof, and (2) a set of existing proofs~\cite{rango} that are likely to share similar proof ideas and can thus assist the LLM in generating a formal proof.

Unlike conventional retrieval methods based on textual similarity between the current subgoal $S$ and the formal lemma statements or subgoal of the proofs in the library, \toolname{} employs a novel \emph{retrieval-with-planning} technique, which leverages the power of LLMs to identify higher-quality lemmas and proofs from the library. Further details about our novel retrieval technique are provided in \Cref{sec:rag}.

\paragraph{LLM-based Proof Generation}
After retrieving lemmas and examples from the library, we are ready to construct a prompt~(details in \Cref{app:prompt}). Our prompt consists of two parts: (i) a fixed system prompt that specifies the Rocq theorem-proving task, and (ii) a user prompt that provides task-specific information for the current subgoal. The user prompt is instantiated from a template and includes the following four components.
\begin{itemize}
    \item First, we provide the current subgoal \(S\) to be proved, together with all formal definitions of the terms that appear in \(S\). We can get these definitions by calling standard APIs of the Rocq prover. This information helps the LLM accurately understand and reason about the current proof task. For example, when instructing the LLM to prove the subgoal shown in \Cref{ex:subgoal}, we include not only the subgoal itself but also the definitions of \texttt{data\_list\_size} and \texttt{dl\_align}.
    \item In addition to the current subgoal, we also provide the retrieved lemmas and proofs.
    \item Moreover, we also provide the LLM with the history of previous failed attempts on the same subgoal, which helps it avoid repeating the same errors. The failure history is maintained per subgoal and updated by the validation-with-reflection module after each failed attempt.
    \item Finally, the prompt instructs the LLM to wrap all generated Rocq tactics with predefined decorators, which facilitates parsing the response from the LLM.
\end{itemize}
After receiving the response from the LLM, our prover can easily parse the LLM-generated formal proof by the design of our prompt. 

\begin{example}
\label{ex:response}
Continuing with \Cref{ex:subgoal}, after performing retrieval and instructing the LLM using the prompt template described above, \toolname{} receives the following response:
\begin{lstlisting}[language=rocq,xleftmargin=10em]
intros l1 l2 pos. induction l1; simpl.
... (* proof below omitted due to irrelevance *)
\end{lstlisting}
\end{example}

\paragraph{Validation with Reflection}
However, LLMs frequently make mistakes when generating tactics, making it necessary to validate their outputs. Traditionally, existing agents only use the proof assistant to validate LLM-generated tactics, and blindly accepts all proofs that are validated to be syntactically correct by the proof assistant. However, an LLM may occasionally misapply a tactic in a way that does not immediately trigger an error. Existing proof agents may still retain such misapplied tactics, causing the search to become trapped in an even harder or invalid branch. Below, we present a concrete example that illustrates this issue.

\begin{example}
\label{ex:induction-wrong}
Continuing with \Cref{ex:response}, all LLM-generated tactics are verified to be syntactically correct by the proof assistant. Intuitively, the proof first introduces the variables \texttt{l1}, \texttt{l2}, and \texttt{pos}, yielding the subgoal shown in \Cref{ex:application}. It then performs induction on \texttt{l1} and simplifies all subgoals generated by the induction. As a result, two subgoals remain, corresponding to the base case and the inductive case; for brevity, we present only the inductive case.

In this case, \texttt{l1} is destructed as \texttt{a :: l1'}. The resulting subgoal (shown below) requires proving that the consequent of \Cref{ex:application} holds for \texttt{l1}, under the inductive hypothesis that the same consequent holds for \texttt{l1'}.


{
\centering
\begin{minipage}[h]{0.9\textwidth}
\begin{lstlisting}[language=rocq,basicstyle=\scriptsize\ttfamily,xleftmargin=0em, escapechar=|]
(* Subgoal for Inductive Case, omit irrelevant details *)
l1'  : list datum
l2   : list datum
a    : datum
pos  : Z
IHl1 : ... /\ dl_align (|\textcolor{red}{pos + data\_list\_size l1'}|) l2
|\hrulefill|
... /\ dl_align (|\textcolor{red}{pos + data\_size a + data\_list\_size l1'}|) l2
\end{lstlisting}
\end{minipage}
\par
}

Although these tactics are validated by the proof assistant, the induction is not effective. Specifically, in the consequent, the first argument of \texttt{dl\_align} differs from the corresponding argument in the inductive hypothesis. This mismatch prevents the inductive hypothesis from being applied, making the subgoal more complicated and even harder to prove.

Instead, the correct approach is to perform induction directly on \texttt{l1} before introducing the remaining variables. This ensures that the inductive hypothesis is sufficiently general and can be applied in the inductive case.
\end{example}

To address this issue, \toolname{} introduces a novel \emph{validation-with-reflection} technique (details in \Cref{sec:reflection}) to filter out potentially misapplied tactics early. Our technique leverages the reasoning capabilities of LLMs during proof validation. This module sequentially examines the generated tactics, checking whether each tactic can be validated by the proof assistant and, more importantly, instructing the LLM to review whether its own generated tactic is potentially misapplied--either ineffective or leading to unprovable subgoals. The output of this module consists of two components:
\begin{itemize}
\item The tactics to be retained for the current iteration, which are both validated by the proof assistant and not flagged as potentially misapplied by the LLM. 
\item A new failure record that updates the failure history. Each record is a triple consisting of (1) the subgoal that failed to be proved, (2) the sequence of tactics attempted in this failure case--either rejected by the proof assistant or flagged as misapplied by the LLM, and (3) the reason for the failure. If the tactics are rejected by the proof assistant, the reason is the corresponding error message; if they are flagged as misapplied, the reason is a concise explanation generated by the LLM.
\end{itemize}
The retained tactics update the set of subgoals, and the proof agent advances to the next iteration with a new set of subgoals.

\begin{example}
Continuing with \Cref{ex:induction-wrong}, our validation-with-reflection module flags these two tactics as misapplied and returns:
\begin{itemize}
\item that no proof should be retained; and
\item a failure record consisting of (1) the subgoal in \Cref{ex:subgoal}, (2) the tactics shown in \Cref{ex:response}, and (3) an LLM-generated explanation stating: “\texttt{The induction is performed without appropriate generalization, making the hypothesis too weak to be applied\ldots}”
\end{itemize}
Since no tactic is retained, the next iteration continues to attempt the same subgoal in \Cref{ex:subgoal}, while the failure history records this unsuccessful induction attempt. In the subsequent iteration, conditioned on this failure record, the LLM avoids repeating the same mistake and successfully generates a complete formal proof of \texttt{dl\_align\_app}.
\end{example}


\newcommand{\rwc}[0]{\texttt{ReflCat}\xspace}
\section{Validation with Reflection}
\label{sec:reflection}

This section presents our novel \emph{validation with reflection} module. The key feature of this module is to use an LLM to self-validate its tactic proposals. This raises two challenges.


\paragraph{Challenges}
First, examining every tactic generated by the LLM is prohibitively costly. Therefore, we adopt a selective strategy that focuses only on tactics that may (1) generate multiple subgoals, or (2) lead to unprovable states (e.g., \texttt{assert}, \texttt{apply}, \texttt{induction}). In contrast, tactics such as \texttt{intros}, \texttt{simpl}, and \texttt{rewrite} are excluded from reflection. We identify the categories of tactics that require reflection by manually inspecting the Rocq tactic index~\cite{CoqTool}. We use \rwc to denote the set of all tactic categories selected for reflection. Due to space limitations, we present \rwc in \Cref{app:refl}.

Second, recall that this module produces failure records that update the failure history. Each record consists of (1) the subgoal that failed to be proved, (2) the sequence of tactics leading to the failure, and (3) the failure reason. Prior proof agents~\cite{copra} record only the subgoal immediately preceding the error and the single tactic that directly triggers it, with the failure reason limited to the error message reported by the proof assistant.

However, failures caused by misapplied tactics often arise from a \emph{sequence} of tactics rather than a single step. For example, applying \texttt{intros} followed by \texttt{induction} may result in insufficient generalization and an unusable inductive hypothesis. In such cases, recording only the subgoal before \texttt{induction} and the \texttt{induction} tactic itself is insufficient; instead, the entire sequence of tactics leading to the failure must be captured. Moreover, because misapplied tactics may not trigger any error from the proof assistant, the failure reason must be explicitly summarized by our system rather than obtained from an error message.

\paragraph{Our Idea}
To address this issue, our module attempts to capture the \emph{problematic sequence of tactics}--that is, the contiguous sequence of tactic applications that leads to a misapplied or ineffective subgoal--using a simple heuristic. \toolname{} maintains the index of the most recent tactic, denoted \(pid\), at which one of the following events occurs: (1) the proof begins, (2) a subgoal is fully proved, or (3) a tactic in \rwc is applied.

When a tactic is flagged as misapplied, the module rolls back to \(pid\) and records a single failure instance consisting of: (1) the subgoal after the first \(pid\) tactics, (2) the sequence of tactics from \(pid\) to the current tactic, and (3) a failure reason summarized by the LLM. Our insight is that proving a subgoal naturally ``closes'' a local proof branch, while tactics in \rwc (e.g., \texttt{assert}, \texttt{apply}, \texttt{induction}) are precisely those that can branch the proof or irreversibly alter the proof structure. Consequently, the sequence of tactics from $pid$ to the current tactic captures a compact yet sufficient context leading to the misapplication. 

\begin{algorithm}[t]
\begin{small}
\caption{Pseudo-code for Validation with Reflection}
\label{fig:pcode}
\DontPrintSemicolon
\KwIn{A list of LLM-generated tactics $T_1\dots T_n$}
\KwOut{The tactics to be saved, and the failure }
\SetKw{Continue}{continue}

$\textit{pid} \gets -1$;\quad $\textit{g}_{\mathrm{pre}} \gets \varnothing$ 

\For{$i \gets 1$ \KwTo $n$}{
  $\textit{g}_{\mathrm{app}}, \textit{err}, \textit{gs}_{\mathrm{new}} \gets \texttt{Execute}(T_i)$\; 
    
  \If{$\textit{err} \neq \varnothing$}{
    \Return $(T_1\dots T_{i-1},\; (\textit{g}_{\mathrm{app}},\; T_i,\; \textit{err}))$\;
  }

  \If{$\textit{gs}_{\mathrm{new}}=\emptyset$} {
    ${\textit{pid}} \gets i$;\quad $\textit{g}_{\mathrm{pre}} \gets \texttt{SubGoals}(0)$\;
    \Continue
  }
  
  \If{$\texttt{Category}(T_i)\in \rwc$}{
    $(\textit{res},\; \textit{summary}) \gets \texttt{Reflection}(\textit{g}_{\mathrm{app}},\; \textit{gs}_{\mathrm{new}},\; T_i)$\;    
    \If{$\textit{res} = \texttt{Misapplied}$}{
      \texttt{Resume($T_{i},\; T_{i-1},\; T_{i-2},\; \dots,\; T_{pid+1}$)}\;
      \Return $(T_1\dots T_{pid},\; (\textit{g}_{\mathrm{pre}},\; T_{\textit{pid}+1}\dots T_i,\; \textit{summary}))$\;
    }
    
    ${\textit{pid}} \gets i$;\quad $\textit{g}_{\mathrm{pre}} \gets \texttt{SubGoals}(0)$ 
  }
}
\Return $(T_1\dots T_n,\; \varnothing)$
\end{small}
\end{algorithm}

\paragraph{Details}
As shown in the pseudo-code in \Cref{fig:pcode}, the module scans the tactics sequentially (Line~2) and maintains two key pieces of state:
(1) \(pid\), the most recent rollback point to be used when a misapplied tactic is detected; and
(2) \(g_{\mathrm{pre}}\), the first unproved subgoal after applying the first \(pid\) tactics.
The subgoal \(g_{\mathrm{pre}}\) is used when constructing failure records, as it captures the subgoal \emph{before} the potentially misapplied tactic sequence begins, providing precise context for diagnosing the failure.

For each tactic \(T_i\), the module first invokes the function \texttt{Execute} (Line~3) with side effect, which executes \(T_i\) in the proof assistant and returns a triple \((g_{\mathrm{app}}, \mathit{err}, gs_{\mathrm{new}})\).
Here, \(g_{\mathrm{app}}\) is the subgoal to which the tactic is applied--by default, the first unproved subgoal prior to execution.
The component \(\mathit{err}\) is the error message returned by the proof assistant, with \(\mathit{err} = \emptyset\) indicating successful execution.
The component \(gs_{\mathrm{new}}\) is the list of new subgoals produced by applying \(T_i\) to \(g_{\mathrm{app}}\).

If the proof assistant reports an error (\(\mathit{err} \neq \emptyset\), Lines~4--5), the module immediately terminates validation: it saves all previously validated tactics \(T_1 \dots T_{i-1}\) and returns a failure record consisting of the current subgoal \(g_{\mathrm{app}}\), the invalid tactic \(T_i\), and the error message \(\mathit{err}\). 

If no new subgoals are generated (\(gs_{\mathrm{new}} = \emptyset\)), the current subgoal is fully proved.
In this case, the module updates \(pid \gets i\) and sets \(g_{\mathrm{pre}}\) to the current first unproved subgoal via \texttt{SubGoals(0)} (Lines 6--8).

If \(T_i\) belongs to \rwc, the module invokes the \texttt{Reflection} subroutine (Lines~9--10), whose details are demonstrated at the end of this section.
This subroutine takes as input the applied subgoal \(g_{\mathrm{app}}\), the newly generated subgoals \(gs_{\mathrm{new}}\), and the tactic \(T_i\), and prompts the LLM to assess whether the tactic is potentially misapplied, i.e., ineffective or leading to unprovable states.
The subroutine returns either \texttt{Accepted} or \texttt{Misapplied}, along with a short LLM-generated explanatory summary.

If \texttt{Misapplied} is returned, the module rolls back the subgoal to the last rollback point \(pid\) and records a failure instance consisting of:
(1) the subgoal \(g_{\mathrm{pre}}\);
(2) the sequence of tactics \(T_{pid+1} \dots T_i\) responsible for the failure; and
(3) the LLM-generated summary explaining why the tactic sequence is misapplied.
Otherwise if \texttt{Accepted} is returned, the module updates \(pid \gets i\) and refreshes \(g_{\mathrm{pre}}\) (Lines 11--14). 

If all tactics are successfully validated, the module returns the full tactic sequence along with an empty failure record (Line~15).

\paragraph{Details of \texttt{Reflection}} Finally, we present the details of the \texttt{Reflection} subroutine. This subroutine performs two checks:
(1) whether all newly generated subgoals in $gs_{\mathrm{new}}$ are provable, i.e., whether any of them is likely to admit counterexamples; and
(2) in the case of the \texttt{induction} tactic, whether the induction is performed on an appropriate variable and whether the inductive hypothesis is sufficiently well generalized.
We treat \texttt{destruct} as a weak form of induction (i.e., induction without an inductive hypothesis) and apply the same checks.
Both checks are carried out by prompting the LLM.

To perform the first check, the system prompt explicitly explains the task, including what it means for a goal to admit a counterexample.
It also provides several in-context learning examples illustrating common cases of unprovable subgoals, such as subgoals with no premises and a contradictory consequent (e.g., \texttt{a < a}).
We verify that these examples do not overlap with any benchmarks used in our experiments.
The user prompt then specifies the concrete task, consisting of all subgoals in $gs_{\mathrm{new}}$ together with all definitions appearing in these subgoals.
We additionally require the LLM to produce a structured, machine-parsable response to facilitate downstream processing. 

To perform the second check, we use a similar prompt with two key differences.
First, the prompt explains common failure modes in induction, such as performing induction on an inappropriate variable or failing to generalize hypotheses properly.
Second, we include the original subgoal $g_{\mathrm{app}}$ in the prompt, allowing the LLM to compare the goal before and after the application of induction and assess whether the induction has been applied appropriately. Due to space limitations, the prompt for both checks are presented in \Cref{app:prompt}.

\section{Retrieval with Planning}
\label{sec:rag}

This section presents our \emph{retrieval with planning} technique. Given the current unproved subgoal, it retrieves relevant lemmas and proofs from the proof library. We begin with a motivating example that highlights a key limitation of subgoal-similarity retrieval, and then describe our method in detail.

\begin{figure}[htbp]
  \centering
  \begin{lstlisting}[language=rocq]
  Theorem approx_scale s m : range1 m (exp2R s * (m / exp2R s)).
  Proof.
    (* Show exp2R s * (m / exp2R s) == m *)
    assert (Em: exp2R s * (m / exp2R s) == m). {
      (* To prove this equality, the key is to use    *)
      (* the lemma mulRCA: x * (y * z) == y * (x * z) *)
      rewrite mulRCA.
      (* Now we need to prove m * (exp2R s / exp2R s) == m, we omit 
      the trivial proof here. *)
      ...
    }
    rewrite Em.
    (* Now it suffices to show range1 m m, which we omit here. *)
    ...
  Qed.
  \end{lstlisting}
  \caption{Motivating Example for Retrieval with Planning.}
  \label{fig:approx-scale}
  \end{figure}

\subsection{Motivating Example}
\label{sec:motivating-example-rag}
Consider the theorem \texttt{approx\_scale} (\Cref{fig:approx-scale}) from the {\textsc Four-Color} project in CoqStoq~\cite{coq-fourcolor}, where:
\begin{itemize}
\item \texttt{range1 m x} is a predicate asserting that a real number $\mathtt{x}\in \mathbb{R}$ lies within the range $[\mathtt{m}, \mathtt{m}+1)$, where $\mathtt{m}\in \mathbb{Z}$ is an integer.
\item \texttt{exp2R s} is a function that takes as input a real number $\mathtt{s}$ and outputs $2^s$. 
\end{itemize}
Overall, the lemma states \texttt{m} dividing and then multiplying by $\mathtt{exp2R\ s}$, whose result is exactly the same as \texttt{m}, still falls into the range $[\mathtt{m}, \mathtt{m}+1)$.

To prove \texttt{approx\_scale}, the crucial step is to establish that the second argument of \texttt{range1} is exactly $\mathtt{m}$, i.e., the assertion \texttt{Em} in \Cref{fig:approx-scale}. Once \texttt{Em} is proved, the goal reduces to \texttt{range1 m m}—namely, $\mathtt{m}\in [\mathtt{m}, \mathtt{m}+1)$—which is immediate. Proving \texttt{Em} requires a small but strategically important algebraic rearrangement: we need to rewrite
\(\texttt{exp2R s * (m / exp2R s)}\)
into a form where $\mathtt{m}$ is isolated and the reciprocal terms are grouped. This step relies on a multiplication-rearrangement lemma, \texttt{mulRCA}, in the proof context.

\begin{figure}[htbp]
  \centering
  \includegraphics[width=0.7\textwidth]{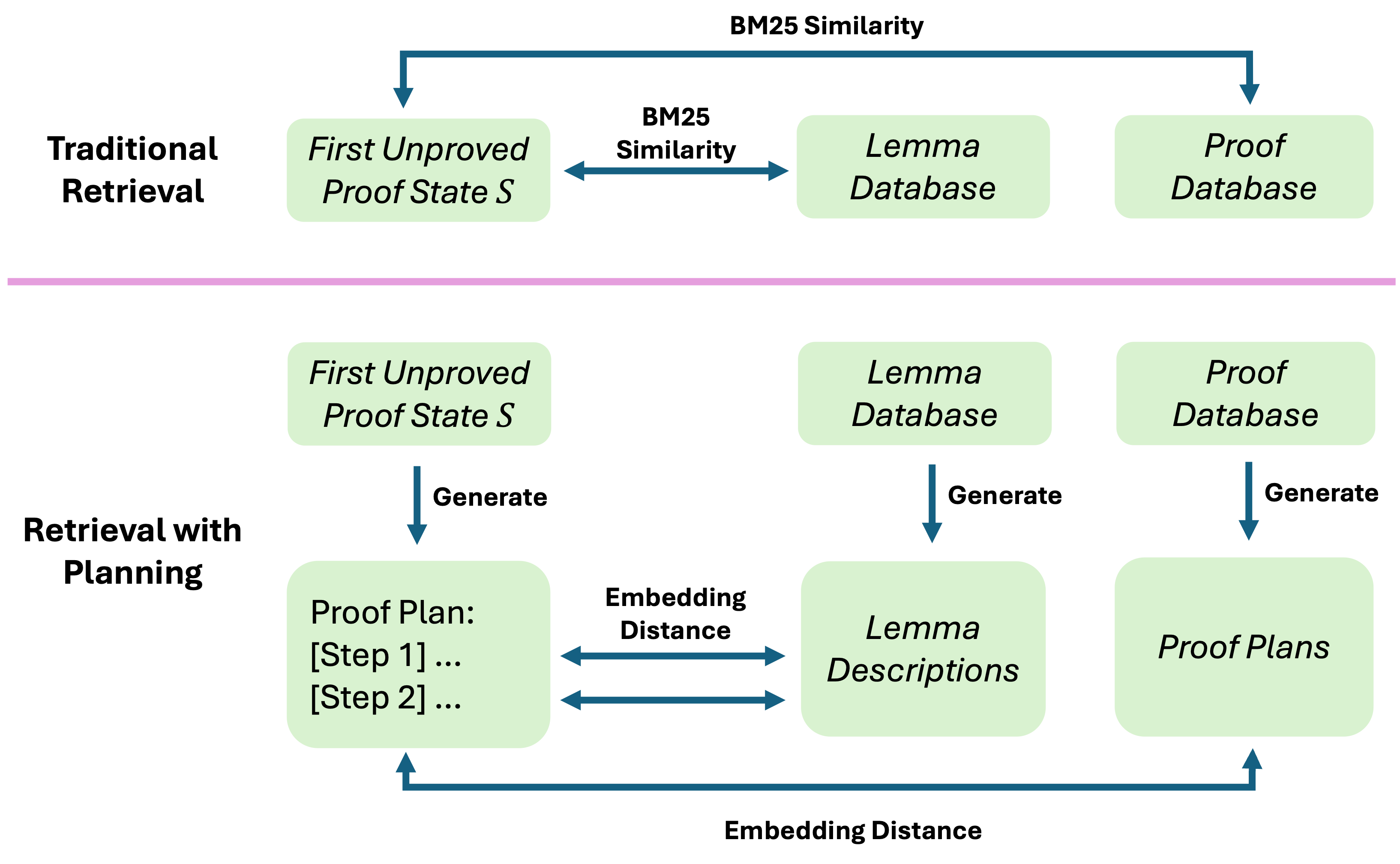}
  \caption{Diagram of Retrieval with Planning. }
  \label{fig:diagram-rag}
\end{figure}

Ideally, the retrieval component should return \texttt{mulRCA}, since it is essential for proving \texttt{Em}. However, standard retrievers (including the one in \Cref{fig:diagram-rag}) rank candidate lemmas by syntactic similarity to the current subgoal. For instance, BM25~\cite{rango}, a widely used baseline, models both the current subgoal and each lemma as bags of terms and scores candidates using term-frequency statistics. While this heuristic often works, it fails in this example because the subgoal is dominated by domain-specific symbols such as \texttt{exp2R} (which appears twice). As a result, BM25 is biased toward lemmas about exponentiation—most of which are irrelevant to the needed algebraic rearrangement. In contrast, \texttt{mulRCA} has little overlap with the subgoal and is therefore ranked low. Consequently, a proof agent relying on BM25 may fail to retrieve \texttt{mulRCA} and cannot complete the proof.

To address this limitation, we introduce \emph{retrieval with planning}, which conditions retrieval on the anticipated proof strategy rather than the raw subgoal. In this example, a plan that explicitly mentions ``rearrange the multiplication to isolate $\mathtt{m}$'' naturally points to \texttt{mulRCA}, even though \texttt{mulRCA} is syntactically dissimilar to the subgoal. \Cref{fig:diagram-rag} presents the overall workflow.

\paragraph{Our Insight}
Our key insight is that LLMs are effective at generating a high-level natural-language proof plan that serves as a good proxy for the structure of the eventual formal proof. However, this plan is expressed in natural language, whereas the available library lemmas are stated in a formal proof language, making direct matching difficult. As illustrated in \Cref{fig:diagram-rag}, we bridge this gap by generating a natural-language description for each lemma and retrieving lemmas based on semantic similarity between each proof-plan step and these descriptions. The semantic similarity is computed in a standard way: we embed each proof-plan step and each lemma description into vectors using an embedding model and rank lemmas by vector similarity (e.g., cosine similarity~\cite{info-retrieval}). Intuitively, semantically related steps and lemmas tend to yield nearby embeddings, and are therefore retrieved together.

\paragraph{Retrieving Proofs} 
Besides retrieving lemmas, prior work~\cite{rango} also retrieves proofs as references to help the LLM construct the current proof. Following the same insight, we generate a natural-language plan for each existing proof and retrieve proofs by embedding-based similarity between the current plan and the stored plans.

\subsection{Details of Retrieval Procedure}
\label{sec:database}

Below, we first present the retrieval procedure for lemmas, and then briefly describe the retrieval procedure for proofs, which follows the same overall structure.

\paragraph{Pre-processing Lemma Databases}
\toolname{} requires a repository of lemmas and proofs. To avoid data leakage, at query time we restrict candidates to the lemmas that are available in the current proof context (i.e., imported and usable at the current location), and retrieve only within this restricted set.

For each lemma, we instruct the LLM to generate a short description that abstracts away syntactic details while preserving the lemma's semantic intent and typical usage~(full details in \Cref{app:prompt}). We then embed this description using a text embedding model that maps text sequences to fixed-dimensional vectors, such as {text-embedding-3-large}~\cite{openai-text-embedding-3-large}, and store the lemma together with its description and embedding vector in the database.
\begin{example}
\label{ex:lemma-description}
Let us continue with \Cref{sec:motivating-example-rag}.
The description of the lemma \texttt{mulRCA} in the proof context generated by the LLM is:
\begin{lstlisting}[basicstyle=\small\ttfamily,frame=single,breaklines=true,escapechar=@,breakindent=0pt,xrightmargin=1em,xleftmargin=0em]
This lemma states that the order of multiplication can be rearranged without changing the result. It can be used to rewrite expressions involving nested multiplications.
\end{lstlisting}
\end{example}

\paragraph{Retrieving Lemmas}
Given an unproved subgoal $S$, we first prompt the LLM to produce a high-level proof plan for $S$. Specifically, the prompt asks for a step-by-step natural-language outline of how to prove the subgoal. To help the LLM generate a better proof plan, in addition to the current subgoal $S$, we also provide the definitions appearing in $S$. To facilitate downstream parsing, we include detailed formatting instructions requiring the plan to be returned as a structured, machine-parsable list (see \Cref{app:prompt} for the full prompt). In response, the LLM outputs a step-by-step proof plan in natural language.
\begin{example}
\label{ex:proof-plan}
Continuing with \Cref{ex:lemma-description},
When we are at the beginning of the proof of theorem \texttt{approx\_scale} in \Cref{fig:approx-scale}, the LLM generates the following proof plan:
\begin{lstlisting}[basicstyle=\small\ttfamily,frame=single,breaklines=true,escapechar=@,breakindent=0pt,xrightmargin=1em,xleftmargin=0em]
<step> Show that the second argument is exactly m </step>
<step> Use a multiplication-rearrangement lemma to rewrite the expression. </step>
(Further steps are omitted for brevity.)
...
\end{lstlisting}
\end{example}

For each generated proof step, \toolname{} embeds the step using the same embedding model and retrieves candidate lemmas by the similarity between the step embedding and the embeddings of lemma descriptions. All retrieved lemmas, along with their descriptions and typical usage, are merged and included in the prompt provided to the LLM.

\begin{example}
\label{ex:retrieval-lemma}
Continuing with \Cref{ex:proof-plan}, note that the second step of the proof plan is semantically similar to the lemma \texttt{mulRCA} in the proof context. Therefore, \toolname{} will retrieve \texttt{mulRCA} as the candidate lemma for the second step.
\end{example}

\paragraph{Retrieval Procedure for Proofs}
The retrieval procedure for proofs is similar to that for lemmas, but differs in two key aspects. First, for each existing proof, we generate a natural-language proof plan using a prompt similar to that described in \Cref{app:prompt}, while omitting any failure history (since none exists) and providing the completed proof. We then embed the resulting proof plan and store each proof together with its plan embedding vector in the database. Second, at query time, we embed the \emph{entire} proof plan for the current unproved state and retrieve those proofs whose plan embeddings are most similar. All retrieved proofs, along with their proof ideas, are included in the prompt provided to the LLM. The prompts for building the proof database are presented in \Cref{app:prompt}.
\section{Evaluation}
\label{sec:evaluation}
This section evaluates the performance of \toolname{}. Specifically, it addresses the following research questions:
\begin{itemize}
\item \textbf{RQ1:} What is the overall performance of \toolname{}?
\item \textbf{RQ2:} How does \toolname{} compare with state-of-the-art proof automation systems?
\item \textbf{RQ3:} What is the effectiveness of each proposed technique, namely validation with reflection (\Cref{sec:reflection}) and retrieval with planning (\Cref{sec:rag})?
\end{itemize}

\paragraph{Benchmark}
We evaluate \toolname{} using CoqStoq~\cite{rango}, a comprehensive benchmark suite for automated proof search in Rocq. CoqStoq is built on Rocq version 8.18 and contains a diverse collection of theorems spanning multiple domains, including compiler verification~\cite{compcert} and formalized mathematics~\cite{coq-fourcolor,coq-coqeal,coq-reglang}.

\paragraph{Implementation Details}
\toolname{} is implemented on Rocq version 8.18 to ensure compatibility with CoqStoq. For CoqHammer, we configure a timeout of 25 seconds and allow up to 64 threads per invocation. All experiments are conducted on a machine equipped with Intel Xeon Gold 6230 CPUs and 503~GB of memory. We set the iteration limit 
$T=25$, allowing at most 25 proof-search iterations per theorem.

Across all experiments, within each experimental setup, we use the same backend language model for retrieval, proof generation, and reflection. For embeddings, we consistently adopt {text-embedding-3-large}~\cite{openai-text-embedding-3-large} as the embedding model. The maximum token limit and temperature for the LLMs are set to their default values.

Recall that \toolname{} requires a repository of lemmas and proofs for retrieval. In our experiments, the retrieval database is the set of theorems in CoqStoq. The database is constructed once offline. At query time, we restrict retrieval to the intersection between this database and the current proof library, ensuring that all retrieved lemmas are available for use in the proof and avoiding potential data leakage. At each retrieval step, we retrieve eight lemmas and eight proofs, and left-clip the final prompt to satisfy the maximum token constraint.

\subsection{RQ1: What is the overall performance of \toolname{}}
\label{sec:rq1}
\paragraph{Setup}
To assess the overall performance of \toolname{}, we evaluate it on the CoqStoq benchmark. Due to budget constraints, we randomly sample 200 theorems from CoqStoq, following prior work~\cite{cobblestone}. We use the state-of-the-art o4-mini model~\cite{o4mini} as the backend language model.

\paragraph{Results}
On the 200 sampled theorems from CoqStoq, \toolname{} successfully proves 138 out of 200 (69\%), demonstrating strong overall performance.

\subsection{RQ2: How does \toolname{} compare to state-of-the-art proof automation systems?}
\label{sec:rq2}

\paragraph{Setup}
We compare \toolname{} against {\sc CobbleStone}, the existing state-of-the-art (SOTA) proof agent. To reduce the cost of LLM invocations, we directly use the detailed evaluation results reported in the open-source repository of {\sc CobbleStone}, without re-running the system. For a fair comparison, we adopt the following configurations for \toolname{}:
\begin{itemize}
    \item {\sc CobbleStone} is configured with GPT-4~\cite{openai-gpt4-technical-report}. Accordingly, we also use GPT-4 as the backend language model for \toolname{}.
    \item {\sc CobbleStone} limits the number of LLM invocations to 20 per theorem. To match this setting, we also restrict \toolname{} to at most 20 LLM invocations per online theorem proving, including those used for retrieval, reflection, and proof generation. Note that each iteration of \toolname{} may invoke multiple LLM invocations--two for retrieval with planning, one for proof generation, and several for validation with reflection. As a result, the prover may terminate before reaching the iteration limit $T=25$. Note that since the database of lemmas and proofs is constructed once, this one-time pre-processing cost is not counted toward the per-theorem LLM-invocation budgets.
    \item {\sc CobbleStone} uses a CoqHammer timeout of 25 seconds and runs on Intel Xeon Gold 6230 hardware. Accordingly, we also configure \toolname{} with a 25-second CoqHammer timeout and conduct all experiments on the same CPU architecture.
\end{itemize}
Furthermore, since {\sc CobbleStone} is built on an earlier version of Rocq (8.12), whereas \toolname{} targets Rocq 8.18, we evaluate both systems on the intersection of the benchmarks used in the {\sc CobbleStone} paper~\cite{cobblestone} and CoqStoq. This results in a total of 222 benchmarks.

\paragraph{Results}
On these 222 benchmarks, {\sc CobbleStone} successfully proves 93 theorems, whereas \toolname{} proves 114, yielding a 22.58\% relative improvement. This performance gap indicates that our approach substantially outperforms the existing state-of-the-art proof agent, with \Cref{ex:induction-wrong} and \Cref{ex:retrieval-lemma} providing concrete illustrations of the effectiveness of our techniques. We make best efforts to align the experimental setup across systems, using the same backend model, the same number of LLM invocations, and the intersection of benchmarks for evaluation. While we cannot guarantee identical configurations in all aspects, we believe the observed performance gap (22.58\% relative improvement) is sufficiently large to outweigh these differences.

Beyond the direct comparison with {\sc CobbleStone}, because we evaluate on the same benchmark set, our results also yield indirect comparisons with the other proof systems~\cite{PALM,ProverBot,rango,Tactician} reported in the {\sc CobbleStone} paper. \Cref{tab:baseline-comparison} summarizes these results. \toolname{} substantially outperforms all prior systems on this benchmark.

\begin{table}[t]
\centering
\caption{Comparison of \toolname{} against baseline proof systems}
\label{tab:baseline-comparison}
\begin{tabular}{l@{\hspace{.5cm}}c@{\hspace{.5cm}}c@{\hspace{.5cm}}c}
\toprule
\textbf{Baseline} & \textbf{\toolname{} (Ours)} & \textbf{Baseline} & \textbf{Rel.\ Improvement} \\
\midrule
{\sc CobbleStone} & 51.35\% & 41.89\% & $+$22.58\% \\
PALM        & 51.35\% & 38.74\% & $+$32.56\% \\
ProverBot9001 & 51.35\% & 17.12\% & $+$200.00\% \\
Rango       & 52.50\% & 36.50\% & $+$43.84\% \\
Tactician   & 52.50\% & 22.50\% & $+$133.33\% \\
\bottomrule
\end{tabular}
\end{table}

\paragraph{Token Efficiency}
In addition to proof success rate, \toolname{} is also substantially more token-efficient than {\sc CobbleStone}. On the 222-benchmark set with the GPT-4 backend, \toolname{} uses an average of \textbf{15.6K tokens} per theorem, compared with {\sc CobbleStone}'s \textbf{48.2K tokens}, making \toolname{} approximately \textbf{3.1$\times$ more token-efficient} while simultaneously solving 22.58\% more theorems.

\subsection{RQ3: Effectiveness of Techniques}
\label{sec:rq3}

\paragraph{Setup}
To assess the effectiveness of each technique in \toolname{}, namely validation with reflection and retrieval with planning, we evaluate \toolname{} with various configurations. We selectively enable or disable the validation-with-reflection module and vary the retrieval strategy between retrieval-with-planning and BM25, the latter being the retrieval method used in prior work~\cite{rango,PALM}. We use the same experimental setup as in \Cref{sec:rq1}.

\begin{table}[t]
\centering
\small
\caption{Results of Different Configurations in Random 200 Samples in CoqStoq}
\label{tab:configurations}
\begin{tabular}{l@{\hspace{0.2cm}}c@{\hspace{0.2cm}}c@{\hspace{0.3cm}}c@{\hspace{0.3cm}}c@{\hspace{0.2cm}}c@{\hspace{0.2cm}}c}
\toprule
\multirow{2}{*}{\bf Configuration} & \multicolumn{4}{c}{\bf Modules} & \multirow{2}{*}{\bf \#Proved} & \multirow{2}{*}{\bf Avg.\ Tokens} \\
\cmidrule(lr){2-5}
 & Hammer & LLM & Refl. & Retrieval &  & \\
\midrule
(C1) Hammer              & \checkmark & $\times$ & $\times$ & $\times$    &  55 {\small$(\uparrow 150.91\%)$} & 24.1K \\
(C2) BM25               & \checkmark & \checkmark & $\times$ & BM25        & 118 {\small$(\uparrow 16.95\%)$}  & 30.0K \\
(C3) Planning           & \checkmark & \checkmark & $\times$ & Planning    & 128 {\small$(\uparrow 7.81\%)$}   & 58.5K \\
(C4) Refl.\ \& BM25     & \checkmark & \checkmark & \checkmark & BM25      & 130 {\small$(\uparrow 6.15\%)$}   & 45.5K \\
(C5) \toolname{}        & \checkmark & \checkmark & \checkmark & Planning  & {\bf 138} & 66.7K \\
\bottomrule
\end{tabular}
\end{table}

\paragraph{Results}
Table~\ref{tab:configurations} presents the result. Each row corresponds to a configuration with different modules enabled or disabled. The ``Modules'' columns indicate whether the configuration uses Hammer, LLM-based proof generation, reflection during validation, and the retrieval strategy employed (none, BM25, or planning). The final column reports \#Proved, the number of problems successfully proved in our random subsamples, and the average token cost per theorem, while the $\uparrow$ percentage indicates the relative improvement of the full system over each configuration.

The full system, \toolname{} (C5), achieves the best performance, proving 138 theorems and outperforming all other configurations. In contrast, the Hammer-only baseline (C1) proves only 55 theorems, highlighting the necessity of LLM-based proof generation in modern proof agents.

Under the same BM25 retrieval strategy, enabling reflection (C4) yields a 10.17\% improvement over the no-reflection setting (C2), demonstrating that reflection-based validation substantially improves proof success. Similarly, under retrieval with planning, reflection (C5) provides a further 7.81\% improvement over its no-reflection counterpart (C3), indicating that reflection consistently contributes additional gains even with stronger retrieval.

Comparing retrieval strategies without validation with reflection, the retrieval with planning (C3) outperforms the BM25 retrieval (C2) by 8.47\%, suggesting that our novel retrieval module is more effective than purely text-similarity based retrieval. When reflection is enabled, the retrieval with planning (C5) still yields an additional 6.15\% improvement over the BM25 retrieval (C4), reinforcing that retrieval with planning remains beneficial even in the presence of validation with reflection. 

Regarding token cost, the full system (C5) uses more tokens per theorem than simpler configurations because of the overhead of plan-based retrieval and reflection. However, this additional cost is worthwhile: the full system (C5) solves 150.91\% more theorems than the no-RAG, no-reflection baseline (C1).

\paragraph{Generalization Across Backend Models}
To assess how well our techniques generalize across language models, we conduct a preliminary experiment using MiniMax-M2.5~\cite{minimax2026m25}, a popular open-source model with 10B active parameters. We randomly sample 100 theorems from CoqStoq and compare \toolname{} (full system) against the BM25-retrieval, no-reflection ablation (matching the C2 vs.\ C5 comparison in \Cref{tab:configurations}). \toolname{} solves 61/100 theorems while the ablation solves 52/100, yielding a \textbf{17.31\% improvement}. This is consistent with the 16.95\% improvement observed under o4-mini (C2 vs.\ C5 in \Cref{tab:configurations}), suggesting that our two proposed techniques generalize across both closed-source and open-source backend models.
 
\paragraph{Statistical Significance}
To further confirm the superiority of \toolname{} over the baselines, we conduct a $p$-test comparing C5 against each configuration. As shown in \Cref{tab:pvalues}, all $p$-values are below 0.05, confirming that the improvements are statistically significant.
 
\begin{table}[t]
\centering
\caption{Statistical significance ($p$-values) of \toolname{} (C5) vs.\ each baseline configuration.}
\label{tab:pvalues}
\begin{tabular}{lc}
\toprule
\textbf{Comparison} & \textbf{$p$-Value} \\
\midrule
vs.\ CobbleStone              & 0.0196 \\
vs.\ (C1) Hammer & 0.0000 \\
vs.\ (C2) BM25 & 0.0002 \\
vs.\ C3 Planning & 0.0075 \\
vs.\ C4 Refl.   & 0.0481 \\
\bottomrule
\end{tabular}
\end{table}

\section{Related Work}
\label{sec:related}

\paragraph{LLM-based proof agents}
Our work is closely related to LLM-based proof agents. A substantial body of prior work has developed agentic systems that combine symbolic provers with LLM-based proof generation~\cite{cobblestone,copra,PALM}. However, as discussed in \Cref{sec:intro}, these systems primarily focus on how to interact with the proof assistant, rather than fully leveraging the reasoning and planning capabilities of LLMs. In contrast, \toolname{} introduces two novel techniques—\emph{validation with reflection} and \emph{retrieval with planning}—that enable LLMs to contribute more effectively to the overall proof-search workflow. Experimental results demonstrate the effectiveness of our approach, achieving a 22.58\% relative improvement in the number of theorems proved over existing state-of-the-art systems.

\paragraph{Language Models for Theorem Proving}
Another related line of research focuses on developing fine-tuned neural theorem provers~\cite{badlur,rango,lean-dojo,lenabel-prover,dsprover,goedel-prover,DSP,ERP}. While these approaches leverage learned models for proof generation, to our knowledge, none of them incorporates the two techniques proposed in \toolname{}, namely validation with reflection and retrieval with planning.

\paragraph{Agents with Planning and Self-reflection}
In the broader AI community, LLM-based agents increasingly combine \emph{planning} and \emph{self-reflection} to improve decision making~\cite{yao2023react,yao2023treeofthoughts,shinn2023reflexion,madaan2023selfrefine}. Adapting these ideas to formal theorem proving, however, is nontrivial. First, the feedback is both sparse and brittle. Many tactic sequences fail abruptly with low-level error messages, while other sequences may be accepted locally yet steer the proof into a potentially unprovable or unnecessarily difficult branch. Second, the available context (e.g., large proof libraries) often exceeds the input budget of LLMs, so planning and reflection must operate under partial information rather than the full set of reusable lemmas and proof patterns. \toolname{} addresses these challenges by introducing two domain-specific mechanisms: \emph{retrieval with planning}, which enables plan-level retrieval of relevant lemmas and proofs under limited context, and \emph{validation with reflection}, which detects and filters potentially misapplied yet syntactically valid tactics before they derail the proof search.



\paragraph{Symbolic provers}
Traditional proof automation in proof assistants relies on symbolic proof search~\cite{coq-hammer,lean-auto,sledgehammer,norman2025canonical}. Some approaches translate proof goals into SMT formulas and then reconstruct proofs as tactics in the proof assistant~\cite{lean-auto,sledgehammer,coq-hammer}, while others employ dedicated proof search over dependent type theory. However, purely symbolic provers suffer from significant scalability limitations. As our experiments indicate, a purely symbolic prover proves only 55/200 theorems in a random sample from CoqStoq, whereas our proof agent proves 138/200 theorems--a 150.91\% relative improvement in the number of theorems proved~(details in \Cref{tab:configurations}).
\section{Conclusion}
\label{sec:conclusion}

This paper advocates for a reasoning-centric approach to automated theorem proving, where LLMs not only generate proof tactics but also strategically plan and critically evaluate their proposals. We address two fundamental limitations in existing LLM-based proof agents: the blind acceptance of syntactically valid but potentially misapplied tactics, and the reliance on syntactic similarity for retrieval rather than strategic alignment.

Our solution introduces two techniques. First, \emph{validation with reflection} enables the LLM to scrutinize its generated tactics through self-reflection, synthesizing failure analyses and regenerating alternatives when potential errors are detected. By selectively targeting tactics whose misuse leads to unprovable subgoals, we maintain efficiency while improving robustness. Second, \emph{retrieval with planning} conditions retrieval on LLM-generated proof plans rather than subgoal similarity. By indexing lemmas through natural-language descriptions and proof examples through strategic plans, we surface knowledge that genuinely aligns with anticipated proof strategies across syntactically diverse theorems.

We implement these innovations in \toolname{}, a proof agent for Rocq evaluated on CoqStoq. Our approach achieves a 22.58\% relative improvement in the number of theorems proved over the previous state-of-the-art, with further studies confirming that both components contribute substantially to overall performance. As reasoning capabilities of language models continue to evolve, positioning reasoning at the heart of automated theorem proving promises increasingly capable systems that effectively complement human expertise in mechanizing complex formal arguments.

 \bibliographystyle{splncs04}
 \bibliography{PL.bib}

@article{coq-fourcolor,
  author  = {Georges Gonthier},
  title   = {Formal Proof---The Four-Color Theorem},
  journal = {Notices of the American Mathematical Society},
  year    = {2008},
  volume  = {55},
  number  = {11},
  pages   = {1382--1393},
  url     = {https://www.ams.org/journals/notices/200811/tx081101382p.pdf}
}

@article{compcert,
  author  = {Xavier Leroy},
  title   = {Formal Verification of a Realistic Compiler},
  journal = {Communications of the ACM},
  year    = {2009},
  volume  = {52},
  number  = {7},
  pages   = {107--115},
  doi     = {10.1145/1538788.1538814},
  url     = {https://xavierleroy.org/publi/compcert-CACM.pdf}
}

@article{coq-hammer,
author = {Czajka, \'{z}Ukasz and Kaliszyk, Cezary},
title = {Hammer for Coq: Automation for Dependent Type Theory},
year = {2018},
issue_date = {June      2018},
publisher = {Springer-Verlag},
address = {Berlin, Heidelberg},
volume = {61},
number = {1–4},
issn = {0168-7433},
url = {https://doi.org/10.1007/s10817-018-9458-4},
doi = {10.1007/s10817-018-9458-4},
journal = {J. Autom. Reason.},
month = jun,
pages = {423–453},
numpages = {31}
}

@InProceedings{sledgehammer,
author="Blanchette, Jasmin Christian
and B{\"o}hme, Sascha
and Paulson, Lawrence C.",
editor="Bj{\o}rner, Nikolaj
and Sofronie-Stokkermans, Viorica",
title="Extending Sledgehammer with SMT Solvers",
booktitle="Automated Deduction -- CADE-23",
year="2011",
publisher="Springer Berlin Heidelberg",
address="Berlin, Heidelberg",
pages="116--130",
isbn="978-3-642-22438-6"
}

@inproceedings{badlur,
  author    = {Emily First and Markus N. Rabe and Talia Ringer and Yuriy Brun},
  title     = {Baldur: Whole-Proof Generation and Repair with Large Language Models},
  booktitle = {Proceedings of the 31st ACM Joint European Software Engineering Conference and Symposium on the Foundations of Software Engineering (ESEC/FSE 2023)},
  pages     = {1229--1241},
  year      = {2023},
  publisher = {ACM},
  address   = {San Francisco, CA, USA},
  doi       = {10.1145/3611643.3616243},
  url       = {https://people.cs.umass.edu/~brun/pubs/pubs/First23fse.pdf}
}

@inproceedings{rango,
  author    = {Kyle Thompson and Nuno Saavedra and Pedro Carrott and Kevin Fisher and Alex Sanchez{-}Stern and Yuriy Brun and Jo{\~a}o F. Ferreira and Sorin Lerner and Emily First},
  title     = {Rango: Adaptive Retrieval-Augmented Proving for Automated Software Verification},
  booktitle = {Proceedings of the 47th International Conference on Software Engineering (ICSE)},
  year      = {2025},
  address   = {Ottawa, Canada},
  month     = apr,
  eprint    = {2412.14063},
  archivePrefix = {arXiv},
  url       = {https://arxiv.org/abs/2412.14063},
  doi       = {10.48550/arXiv.2412.14063},
  note      = {To appear; ICSE 2025}
}

@article{cobblestone,
  author       = {Saketh Ram Kasibatla and Arpan Agarwal and Yuriy Brun and Sorin Lerner and Talia Ringer and Emily First},
  title        = {Cobblestone: A Divide-and-Conquer Approach for Automating Formal Verification},
  year         = {2025},
  eprint       = {2410.19940},
  archivePrefix= {arXiv},
  primaryClass = {cs.LO},
  url          = {https://arxiv.org/abs/2410.19940},
  doi          = {10.48550/arXiv.2410.19940},
  note         = {arXiv v3 (Aug 2025)}
}

@inproceedings{lean-dojo,
  author    = {Kaiyu Yang and Aidan M. Swope and Alex Gu and Rahul Chalamala and Peiyang Song and Shixing Yu and Saad Godil and Ryan J. Prenger and Anima Anandkumar},
  title     = {LeanDojo: Theorem Proving with Retrieval-Augmented Language Models},
  booktitle = {Advances in Neural Information Processing Systems 36 (NeurIPS 2023), Datasets and Benchmarks Track},
  year      = {2023},
  url       = {https://proceedings.neurips.cc/paper_files/paper/2023/file/4441469427094f8873d0fecb0c4e1cee-Paper-Datasets_and_Benchmarks.pdf},
  note      = {Includes Lean datasets/benchmarks (Lean 3/4) for training and evaluation}
}

@inproceedings{copra,
  author    = {Amitayush Thakur and George Tsoukalas and Yeming Wen and Jimmy Xin and Swarat Chaudhuri},
  title     = {An In-Context Learning Agent for Formal Theorem-Proving},
  booktitle = {Proceedings of the 1st Conference on Language Modeling (COLM 2024)},
  year      = {2024},
  month     = oct,
  address   = {Philadelphia, PA, USA},
  url       = {https://openreview.net/forum?id=V7HRrxXUhN},
  note      = {COLM 2024, Oct 7--9, 2024}
}

@inproceedings{yao2023react,
  author    = {Shunyu Yao and Jeffrey Zhao and Dian Yu and Nan Du and Izhak Shafran and Karthik R. Narasimhan and Yuan Cao},
  title     = {ReAct: Synergizing Reasoning and Acting in Language Models},
  booktitle = {The Eleventh International Conference on Learning Representations (ICLR 2023), Kigali, Rwanda, May 1--5, 2023},
  publisher = {OpenReview.net},
  year      = {2023},
  url       = {https://openreview.net/forum?id=WE_vluYUL-X}
}

@online{openai-text-embedding-3-large,
  author       = {{OpenAI}},
  title        = {{text-embedding-3-large}},
  date         = {2024-01-25},
  organization = {OpenAI},
  url          = {https://platform.openai.com/docs/models/text-embedding-3-large},
  urldate      = {2026-01-28},
  note         = {OpenAI API embedding model documentation; model released 2024-01-25.},
}

@online{o4mini,
  author    = {{OpenAI}},
  title     = {o4-mini: latest reasoning model in the OpenAI o-series},
  year      = {2025},
  date      = {2025-04-16},
  url       = {https://platform.openai.com/docs/models/o4-mini},
  urldate   = {2026-01-28}
}

@report{openai-gpt4-technical-report,
  author       = {{OpenAI}},
  title        = {GPT-4 Technical Report},
  type         = {Technical Report},
  institution  = {OpenAI},
  date         = {2023-03-27},
  eprint       = {2303.08774},
  eprinttype   = {arxiv},
  eprintclass  = {cs.CL},
  url          = {https://arxiv.org/abs/2303.08774},
  urldate      = {2026-01-27},
}

@misc{dsprover,
      title={DeepSeek-Prover-V2: Advancing Formal Mathematical Reasoning via Reinforcement Learning for Subgoal Decomposition}, 
      author={Z. Z. Ren and Zhihong Shao and Junxiao Song and Huajian Xin and Haocheng Wang and Wanjia Zhao and Liyue Zhang and Zhe Fu and Qihao Zhu and Dejian Yang and Z. F. Wu and Zhibin Gou and Shirong Ma and Hongxuan Tang and Yuxuan Liu and Wenjun Gao and Daya Guo and Chong Ruan},
      year={2025},
      eprint={2504.21801},
      archivePrefix={arXiv},
      primaryClass={cs.CL},
      url={https://arxiv.org/abs/2504.21801}, 
}

@misc{lenabel-prover,
  author       = {Zhang, Jingyuan and Wang, Qi and Ji, Xingguang and Liu, Yahui and Yue, Yang and Zhang, Fuzheng and Zhang, Di and Zhou, Guorui and Gai, Kun},
  title        = {Leanabell-Prover: Posttraining Scaling in Formal Reasoning},
  year         = {2025},
  url          = {https://arxiv.org/abs/2504.06122}
}

@article{goedel-prover,
  author       = {Lin, Yong and Tang, Shange and Lyu, Bohan and Ziran Yang and Jui-Hui Chung and Haoyu Zhao and Lai Jiang and Yihan Geng and Jiawei Ge and Jingruo Sun and Jiayun Wu and Jiri Gesi and Ximing Lu and David Acuna and Kaiyu Yang and Hongzhou Lin and Yejin Choi and Danqi Chen and Sanjeev Arora and Chi Jin},
  title        = {Goedel-Prover V2: Scaling Formal Theorem Proving with Scaffolded Data Synthesis and Self-Correction},
  journal      = {arXiv preprint arXiv:2508.03613},
  year         = {2025},
  note         = {Version V2, improved pipeline and SOTA results},
  url          = {https://arxiv.org/abs/2508.03613}
}

@article{PALM,
  author  = {Aakanksha Chowdhery and Sharan Narang and Jacob Devlin and Maarten Bosma and Gaurav Mishra and Adam Roberts and Paul Barham and Hyung Won Chung and Charles Sutton and Sebastian Gehrmann and Parker Schuh and Kensen Shi and Sasha Tsvyashchenko and Joshua Maynez and Abhishek Rao and Parker Barnes and Yi Tay and Noam Shazeer and Vinodkumar Prabhakaran and Emily Reif and Nan Du and Ben Hutchinson and Reiner Pope and James Bradbury and Guy Gur-Ari and Pengcheng Yin and Toju Duke and Anselm Levskaya and Sanjay Ghemawat and Sunipa Dev and Henryk Michalewski and Xavier Garcia and Vedant Misra and Kevin Robinson and Liam Fedus and Denny Zhou and Daphne Ippolito and David Luan and Hyeontaek Lim and Barret Zoph and Alexander Spiridonov and Ryan Sepassi and David Dohan and Shivani Agrawal and Mark Omernick and Andrew M. Dai and Thanumalayan Sankaranarayana Pillai and Marie Pellat and Aitor Lewkowycz and Erica Moreira and Rewon Child and Oleksandr Polozov and Katherine Lee and Zongwei Zhou and Xuezhi Wang and Brennan Saeta and Mark Diaz and Orhan Firat and Michele Catasta and Jason Wei and Kathy Meier{-}Hellstern and Douglas Eck and Jeff Dean and Slav Petrov and Noah Fiedel},
  title   = {PaLM: Scaling Language Modeling with Pathways},
  journal = {Journal of Machine Learning Research},
  volume  = {24},
  number  = {240},
  pages   = {1--113},
  year    = {2023},
  url     = {https://jmlr.org/papers/v24/22-1144.html}
}

@InProceedings{lean-auto,
author="Qian, Yicheng
and Clune, Joshua
and Barrett, Clark
and Avigad, Jeremy",
editor="Piskac, Ruzica
and Rakamari{\'{c}}, Zvonimir",
title="Lean-Auto: An Interface Between Lean 4 and Automated Theorem Provers",
booktitle="Computer Aided Verification",
year="2025",
publisher="Springer Nature Switzerland",
address="Cham",
pages="175--196",
isbn="978-3-031-98682-6"
}

@article{coq-reglang,
author = {Doczkal, Christian and Smolka, Gert},
title = {Regular Language Representations in the Constructive Type Theory of Coq},
year = {2018},
issue_date = {June      2018},
publisher = {Springer-Verlag},
address = {Berlin, Heidelberg},
volume = {61},
number = {1–4},
issn = {0168-7433},
url = {https://doi.org/10.1007/s10817-018-9460-x},
doi = {10.1007/s10817-018-9460-x},
journal = {J. Autom. Reason.},
month = jun,
pages = {521–553},
numpages = {33}
}

@InProceedings{coq-coqeal,
author="Cohen, Cyril
and M{\"o}rtberg, Anders",
editor="Klein, Gerwin
and Gamboa, Ruben",
title="A Coq Formalization of Finitely Presented Modules",
booktitle="Interactive Theorem Proving",
year="2014",
publisher="Springer International Publishing",
address="Cham",
pages="193--208",
isbn="978-3-319-08970-6"
}

@book{CoqTool,
author = {Bertot, Yves and Casteran, Pierre},
title = {Interactive Theorem Proving and Program Development},
year = {2004},
isbn = {3540208542},
publisher = {SpringerVerlag}
}

@inproceedings{DSP,
  title     = {Draft, Sketch, and Prove: Guiding Formal Theorem Provers with Informal Proofs},
  author    = {Jiang, Albert Qiaochu and Welleck, Sean and Zhou, Jin Peng and Li, Wenda and Liu, Jiacheng and Jamnik, Mateja and Lacroix, Timoth{\'e}e and Wu, Yuhuai and Lample, Guillaume},
  booktitle = {International Conference on Learning Representations},
  year      = {2023},
  url       = {https://arxiv.org/abs/2210.12283}
}

@inproceedings{ERP,
  title     = {ProofAug: Efficient Neural Theorem Proving via Fine-grained Proof Structure Analysis},
  author    = {Liu, Haoxiong and Sun, Jiacheng and Li, Zhenguo and Yao, Andrew C.},
  booktitle = {Proceedings of the 42nd International Conference on Machine Learning},
  pages     = {39568--39586},
  year      = {2025},
  editor    = {Singh, Aarti and Fazel, Maryam and Hsu, Daniel and Lacoste-Julien, Simon and Berkenkamp, Felix and Maharaj, Tegan and Wagstaff, Kiri and Zhu, Jerry},
  volume    = {267},
  series    = {Proceedings of Machine Learning Research},
  month     = {13--19 Jul},
  publisher = {PMLR},
  url       = {https://proceedings.mlr.press/v267/liu25bp.html}
}

@inproceedings{shinn2023reflexion,
author = {Shinn, Noah and Cassano, Federico and Gopinath, Ashwin and Narasimhan, Karthik and Yao, Shunyu},
title = {Reflexion: language agents with verbal reinforcement learning},
year = {2023},
publisher = {Curran Associates Inc.},
address = {Red Hook, NY, USA},
booktitle = {Proceedings of the 37th International Conference on Neural Information Processing Systems},
articleno = {377},
numpages = {19},
location = {New Orleans, LA, USA},
series = {NIPS '23}
}

@inproceedings{madaan2023selfrefine,
author = {Madaan, Aman and Tandon, Niket and Gupta, Prakhar and Hallinan, Skyler and Gao, Luyu and Wiegreffe, Sarah and Alon, Uri and Dziri, Nouha and Prabhumoye, Shrimai and Yang, Yiming and Gupta, Shashank and Majumder, Bodhisattwa Prasad and Hermann, Katherine and Welleck, Sean and Yazdanbakhsh, Amir and Clark, Peter},
title = {SELF-REFINE: iterative refinement with self-feedback},
year = {2023},
publisher = {Curran Associates Inc.},
address = {Red Hook, NY, USA},
booktitle = {Proceedings of the 37th International Conference on Neural Information Processing Systems},
articleno = {2019},
numpages = {61},
location = {New Orleans, LA, USA},
series = {NIPS '23}
}

@inproceedings{yao2023treeofthoughts,
author = {Yao, Shunyu and Yu, Dian and Zhao, Jeffrey and Shafran, Izhak and Griffiths, Thomas L. and Cao, Yuan and Narasimhan, Karthik},
title = {Tree of thoughts: deliberate problem solving with large language models},
year = {2023},
publisher = {Curran Associates Inc.},
address = {Red Hook, NY, USA},
booktitle = {Proceedings of the 37th International Conference on Neural Information Processing Systems},
articleno = {517},
numpages = {14},
location = {New Orleans, LA, USA},
series = {NIPS '23}
}

@book{info-retrieval,
  title     = {Introduction to Information Retrieval},
  author    = {Manning, Christopher D. and Raghavan, Prabhakar and Sch{\"u}tze, Hinrich},
  year      = {2008},
  publisher = {Cambridge University Press},
  address   = {Cambridge, UK},
  chapter   = {6}
}

@inproceedings{ProverBot,
author = {Sanchez-Stern, Alex and Alhessi, Yousef and Saul, Lawrence and Lerner, Sorin},
title = {Generating correctness proofs with neural networks},
year = {2020},
isbn = {9781450379960},
publisher = {Association for Computing Machinery},
address = {New York, NY, USA},
url = {https://doi.org/10.1145/3394450.3397466},
doi = {10.1145/3394450.3397466},
booktitle = {Proceedings of the 4th ACM SIGPLAN International Workshop on Machine Learning and Programming Languages},
pages = {1–10},
numpages = {10},
location = {London, UK},
series = {MAPL 2020}
}

@inproceedings{Tactician,
author = {Blaauwbroek, Lasse and Urban, Josef and Geuvers, Herman},
title = {The Tactician: A Seamless, Interactive Tactic Learner and Prover for Coq},
year = {2020},
isbn = {978-3-030-53517-9},
publisher = {Springer-Verlag},
address = {Berlin, Heidelberg},
url = {https://doi.org/10.1007/978-3-030-53518-6\_17},
doi = {10.1007/978-3-030-53518-6\_17},
booktitle = {Intelligent Computer Mathematics: 13th International Conference, CICM 2020, Bertinoro, Italy, July 26–31, 2020, Proceedings},
pages = {271–277},
numpages = {7},
location = {Bertinoro, Italy}
}

@misc{minimax2026m25,
  author = {{MiniMax}},
  title  = {{MiniMax-M2.5}: Built for Real-World Productivity},
  year   = {2026},
  howpublished = {\url{https://www.minimax.io/news/minimax-m25}},
  note   = {Accessed: 2026}
}

@inproceedings{norman2025canonical,
  author    = {Chase Norman and Jeremy Avigad},
  title     = {Canonical for Automated Theorem Proving in Lean},
  booktitle = {16th International Conference on Interactive Theorem Proving (ITP 2025)},
  series    = {Leibniz International Proceedings in Informatics (LIPIcs)},
  volume    = {352},
  pages     = {14:1--14:20},
  year      = {2025},
  publisher = {Schloss Dagstuhl--Leibniz-Zentrum f{\"u}r Informatik},
  doi       = {10.4230/LIPIcs.ITP.2025.14},
  url       = {https://drops.dagstuhl.de/entities/document/10.4230/LIPIcs.ITP.2025.14}
}
\appendix
\section{Omitted Prompts}
\label{app:prompt}
\subsection{Prompts for LLM-based Proof Generation}
\begin{lstlisting}[basicstyle=\small\ttfamily,frame=single,breaklines=true,escapechar=@,breakindent=0pt,xrightmargin=1em,xleftmargin=0em]
  @\textbf{{[System Prompt]}}@
  You are an expert in Rocq theorem proving. Your task is to generate a sequence of tactics to prove the given subgoal. You will be provided with: (i) the current subgoal, and (ii) contextual information including relevant definitions, illustrative examples, applicable lemmas, and the history of previous proof attempts. Based on this information, generate a valid proof script that will be accepted by the Rocq proof assistant.
  @\textbf{{[User Prompt]}}@
  ### subgoal to be Solved ...
  ### Definitions ...
  ### Examples ...
  ### Lemmas ...
  ### Failure History ...
  You need to wrap generated tactics with <coq> and </coq>.
  \end{lstlisting}
\subsection{Prompts for Evaluating the Provability of Subgoals}
\begin{lstlisting}[basicstyle=\small\ttfamily,frame=single,breaklines=true,escapechar=@,breakindent=0pt,xrightmargin=1em,xleftmargin=0em]
[System Prompts]
# Rocq Provability Evaluation Module

You are an expert Rocq proof assistant specializing in evaluating the provability of proof goals. Your role is to analyze whether the current proof goals are likely to be provable or if they contain logical contradictions, false assumptions, or other issues that would make them impossible to prove.

## Your Task

Given:
- **Current Goals**: The obtained proof subgoal(s) after applying the tactic
- **Relevant Definitions**: All definitions from the codebase that are relevant to understanding the goal structure

You must evaluate whether the current goals are provable, identifying any issues that would prevent successful completion of the proof.

## Common Issues to Detect

### 1. Contradictory Hypotheses
When the hypotheses contain logical contradictions that make the goal vacuously true but practically unprovable.

**Example of Unprovable Goal:**
- Hypotheses: `n : nat`, `H1 : n = 0`, `H2 : n = 1`
- Goal: `n + 1 = 2`
- Problem: The hypotheses contradict each other (n cannot be both 0 and 1), making this state impossible to reach in a valid proof

**What to check:**
- Look for hypotheses that assert contradictory facts
- Check for equality assumptions that conflict with each other
- Identify impossible combinations of conditions

### 2. Missing or Insufficient Hypotheses
When the goal requires information that is not available in the hypotheses or context.

**Example of Unprovable Goal:**
- Hypotheses: `n : nat`
- Goal: `n > 10`
- Problem: There's no information about n that would allow proving it's greater than 10

**What to check:**
- Does the goal make claims that can be derived from the hypotheses?
- Are all necessary facts about variables present in the context?
- Would the goal require additional axioms or lemmas not available?

### 3. Overly Strong or Unprovable Statements
When the goal makes a claim that is mathematically false or requires axioms not available.

**Example of Unprovable Goal:**
- Goal: `forall n : nat, exists m : nat, m > n /\ m < n`
- Problem: This is logically impossible (no number can be both greater than and less than another number)

**What to check:**
- Is the goal mathematically/logically valid?
- Does it require non-constructive reasoning in a constructive logic?
- Are there claims that would require additional axioms (e.g., excluded middle)?

## Output Format

You MUST respond with the following structured format:

```markdown
### Analysis
[Your detailed analysis of the current proof goals, explaining what you observe and any potential issues]

### Decision
[PROVABLE, UNPROVABLE, or UNCERTAIN]

- **PROVABLE**: The goals appear to be logically valid and provable with available tactics and lemmas
- **UNPROVABLE**: The goals contain clear contradictions or impossible requirements
- **UNCERTAIN**: Unable to determine with confidence; may require specialized knowledge or techniques

### Reason
[Brief explanation of your decision, highlighting the key factors]

### Suggestion
[If UNPROVABLE, provide suggestions on how to fix the issue, such as:
- What hypotheses need to be corrected
- What additional lemmas might be needed
You may include Rocq code blocks with ```rocq for concrete suggestions.
If PROVABLE or UNCERTAIN, output "N/A"]
```

## Important Guidelines

1. Focus on logical validity and provability, not on finding the optimal proof strategy
2. Be conservative: if you're not sure whether something is provable, mark it as UNCERTAIN rather than UNPROVABLE
3. Consider that the proof may require advanced techniques you're not aware of - don't mark something UNPROVABLE unless you're confident
4. Check for contradictions carefully - subtle contradictions can make goals unprovable
5. Consider the constructive nature of Rocq's logic - some classically true statements may not be constructively provable
6. Always follow the exact output format for parseability
7. Provide actionable suggestions when marking goals as UNPROVABLE
8. Remember that "difficult" does not mean "unprovable"

[User Prompts]
### Current Goals ...
### Relevant Definitions ...
\end{lstlisting}

\subsection{Prompts for Induction Schema Evaluation}
\begin{lstlisting}[basicstyle=\small\ttfamily,frame=single,breaklines=true,escapechar=@,breakindent=0pt,xrightmargin=1em,xleftmargin=0em]
[System Prompt]
# Rocq Induction Evaluation Module

You are an expert Rocq proof assistant specializing in evaluating induction strategies. Your role is to analyze whether an induction tactic was applied reasonably and effectively. Please treat destruct as a weaker form of induction with no inductive hypothesis.

## Your Task

Given:
1. **Goal Before Induction**: The original proof goal
2. **Goal After Induction**: The resulting subgoals after applying induction
3. **Induction Strategies**: The specific tactic(s) used for induction strategy (e.g., intro+induction pattern)
4. **Relevant Definitions**: Key definitions from the codebase that are relevant to understanding the goal structure and determining the appropriate induction strategy

You must evaluate whether the induction was performed reasonably, particularly checking if variables were over-bound before induction. Use the relevant definitions to understand the recursive structure of functions and datatypes involved in the goal.

## Common Issues to Detect

### 1. Over-binding Variables
When proving statements like `forall n m, P n m`, if you introduce both `n` and `m` into the context before applying induction on `n`, the induction hypothesis will only apply to the specific `m` already in the context. This makes the induction hypothesis too weak, as it doesn't generalize over all possible values of `m`.

**Example of Unreasonable Induction:**
- Goal: `forall n m, n + m = m + n`
- UNREASONABLE Tactic: `intros n m. induction n.` (over-bound m before induction)
- Problem: The induction hypothesis becomes `n + m = m + n -> S n + m = m + S n` for the specific `m` in context, rather than `forall m, n + m = m + n -> forall m, S n + m = m + S n`. This makes the proof impossible or much harder.

**Reasonable Version:**
- Avoid introducing m before induction: `intros n. induction n. intros m.`
- Or use: `induction n; intros m.`
- If already introduced: `intros n m. revert m. induction n. intros m.`

### 2. Wrong Variable for Induction
Choosing to induct on a variable that doesn't appear in the toplevel match.

**Example of Unreasonable Induction:**
- Relevant Definitions: 
  ```coq
  Fixpoint plus (n m : nat) : nat :=
    match n with
    | O => m
    | S p => S (plus p m)
    end.
  ```
- Goal: `forall n m, n + m = m + n`
- UNREASONABLE Tactic: `induction m.` (note that the recursive definition of + matches on `n` instead of `m` in the toplevel)
- Problem: Inducting on `m` does not simplify the `match` expression in the definition of `+`, which matches on `n`. This makes the proof much harder.
**Reasonable Version:**
- Should induct on n: `induction n.`

#### Hint: **Induction Should Choose the Variable with Toplevel Match**

**Key Principle:** When a goal involves a function, induction should be performed on the variable that appears in the **toplevel match** of that function's definition. This aligns the induction with the recursive structure of the function.

**How to identify the correct variable:**
1. Look at the relevant function definitions in the goal
2. Find the **toplevel match** statement (ignore nested matches)
3. The variable being matched at the toplevel is the one you should induct on

**Example 1 - Simple Case:**
```coq
Fixpoint plus (n m : nat) : nat :=
  match n with  (* toplevel match on n *)
  | O => m
  | S p => S (plus p m)
  end.
```
- Toplevel match is on `n`
- **Correct induction:** `induction n`
- **Incorrect induction:** `induction m` (m does not appear in toplevel match)

**Example 2 - List Append:**
```coq
Fixpoint app (A : Type) (xs ys : list A) : list A :=
  match xs with  (* toplevel match on xs *)
  | nil => ys
  | cons x xs' => cons x (app A xs' ys)
  end.
```
- Toplevel match is on `xs`
- **Correct induction:** `induction xs`
- **Incorrect induction:** `induction ys` (ys does not appear in toplevel match)

**Example 3 - Nested Matches:**
```coq
Fixpoint some_function (n m : nat) : nat :=
  match n with  (* toplevel match on n - this determines induction variable! *)
  | O => match m with  (* nested match - ignore for induction choice *)
         | O => 0
         | S m' => m
         end
  | S n' => S (some_function n' m)
  end.
```
- Toplevel match is on `n` (the nested match on `m` is irrelevant)
- **Correct induction:** `induction n`
- **Incorrect induction:** `induction m` (m only appears in nested match)



### 3. Over-binding Multiple Variables
When multiple variables appear after the induction variable, over-binding all of them before induction makes the induction hypothesis too weak.

**Example of Unreasonable Induction:**
- Goal: `forall n m k, n + (m + k) = (n + m) + k`
- UNREASONABLE Tactic: `intros n m k. induction n.` (over-bound m and k)
- Problem: The induction hypothesis is too weak - it only proves the property for the specific m and k in context, not for all m and k

**Reasonable Version:**
- Should avoid over-binding m and k: `intros n. induction n. intros m k.`
- Or use: `induction n; intros m k.`

**Another Example:**
- Goal: `forall xs ys zs, app xs (app ys zs) = app (app xs ys) zs`
- UNREASONABLE Tactic: `intros xs ys zs. induction xs.` (over-bound ys and zs)
- Problem: The induction hypothesis won't be strong enough to handle arbitrary lists ys and zs

**Reasonable Version:**
- Should use: `induction xs; intros ys zs.` or `intros xs. induction xs. intros ys zs.`


## Output Format

You MUST respond with the following structured format:

```markdown
### Analysis
[Your detailed analysis of the induction strategy, explaining what was done and why it may or may not be reasonable]

### Decision
[REASONABLE or UNREASONABLE]

### Reason
[Brief explanation of your decision]

### Suggestion
[If UNREASONABLE, output in the following form:
```coq
[the reasonable version of the tactic sequence]
```
If REASONABLE, output "N/A"]
```

## Important Guidelines

1. Focus on whether the induction hypothesis will be strong enough to complete the proof
2. Check if universally quantified variables that appear after the induction variable were over-bound (introduced before induction when they shouldn't be)
3. Consider the structure of the goal and what the induction hypothesis needs to prove
4. Be specific in your suggestions - provide actual Rocq tactic syntax
5. Always follow the exact output format for parseability
6. Look at `Relevant Definitions` why checking `Wrong Variable for Induction`
7. Think carefully for a good answer.
8. Wrap the code in a ```rocq code block in the suggestion part.
9. Provide ONLY ONE better version of tactic sequence.

[User prompts]
### Goal Before Induction ...
### Goal After Induction ...
### Induction Strategies ...
### Relevant Definitions ...
\end{lstlisting}

\subsection{Prompts for Building Retrieval Database for Lemmas}
\begin{lstlisting}[basicstyle=\small\ttfamily,frame=single,breaklines=true,escapechar=@,breakindent=0pt,xrightmargin=1em,xleftmargin=0em]
  @\textbf{{[System Prompt]}}@
  You are an expert in Rocq theorem proving. Your task is to generate a concise natural-language description for the given lemma. You will be provided with the lemma statement and the definitions of all terms in the lemma statement. Your description should explain: (1) what the lemma states, and (2) the scenarios in which this lemma is typically used.
  @\textbf{{[User Prompt]}}@
  ### Lemma Statement ...
  ### Definitions ...
  \end{lstlisting}

\subsection{Prompts for Retrieving Lemmas and Proofs}

\begin{lstlisting}[basicstyle=\small\ttfamily,frame=single,breaklines=true,escapechar=@,breakindent=0pt,xrightmargin=1em,xleftmargin=0em]
  @\textbf{{[System Prompt]}}@
  You are an expert in Rocq theorem proving. Your task is to generate a concise natural-language step-by-step proof plan for the given subgoal. You will be provided with the subgoal, as well as the definitions of all terms in the subgoal. Your plan should be in natural language. Please generate in the following format:
  <step> The step 1 description </step>
  <step> The step 2 description </step>
  ...
  @\textbf{{[User Prompt]}}@
  ### Subgoal ...
  ### Definitions ...
  \end{lstlisting}

\subsection{Prompts for Building Retrieval Database for Proofs}

\begin{lstlisting}[basicstyle=\small\ttfamily,frame=single,breaklines=true,escapechar=@,breakindent=0pt,xrightmargin=1em,xleftmargin=0em]
@\textbf{{[System Prompt]}}@
You are an expert in Rocq theorem proving. Your task is to generate a concise natural-language step-by-step proof plan for the given subgoal. You will be provided with the subgoal, the formal proof to this subgoal, and the definitions of all terms in the subgoal and the proof. Your plan should be in natural language. Please generate in the following format:
<step> The step 1 description </step>
<step> The step 2 description </step>
...
@\textbf{{[User Prompt]}}@
### Subgoal ...
### Formal Proof ...
### Definitions ...
\end{lstlisting}


\section{Details in Reflection}
\label{app:refl}
Below, we present the complete list of \rwc. The set \rwc consists of the following categories of tactics:
\begin{itemize}
    \item Introducing auxiliary subgoals: \texttt{assert}, \texttt{have}, \texttt{pose}.
    \item Applying lemmas: \texttt{apply}, \texttt{eapply}.
    \item Choosing a proof branch: \texttt{left}, \texttt{right}.
    \item Induction-related tactics: \texttt{induction}, \texttt{destruct}, \texttt{case}, \texttt{elim}.
\end{itemize}

\end{document}